\documentclass[twocolumn,superscriptaddress,showpacs,prl,aps,amsmath,amssymb,nofootinbib]{revtex4-1}
\usepackage{natbib}
\usepackage{graphicx,color}
\usepackage{amsmath,amssymb}
\usepackage{verbatim}
\usepackage{float}
\usepackage{wasysym}
\usepackage{amssymb,graphicx}
\usepackage{epsfig}
\usepackage{psfrag}
\usepackage{dsfont}
\usepackage{amsfonts}
\usepackage{mathrsfs}
\usepackage{multirow}
\usepackage{times}
\usepackage{bm}
\usepackage{hyperref}
\usepackage{xspace}
\hypersetup{
  colorlinks=true,        
  linkcolor=blue,         
  citecolor=cyan,         
}
\usepackage{pifont}
\usepackage{multirow}
\usepackage{verbatim}

\graphicspath{{./}}

\newcommand{\beq}{\begin{equation}} 
\newcommand{\eeq}{\end{equation}} 
\newcommand{\beqn}{\begin{eqnarray}} 
\newcommand{\eeqn}{\end{eqnarray}}

\newcommand{\zD}{{\raise1.0ex\hbox{${}^{\ \circ}$}}\!\!\!\!\!D}
\newcommand{\alone}{{\raise0.5ex\hbox{${}^{\ 1}$}}\!\!\!\!\alpha}

\newcommand{\nalam}{\mathrel{\raise0.9ex\hbox{$^\lambda$}\mkern-14mu
\lower0.0ex\hbox{$\nabla$}}}

\newcommand{\zeroD}{{\raise1.0ex\hbox{${}^{\ \circ}$}}\!\!\!\!\!D}

\newcommand{\zLap}{{\raise1.0ex\hbox{${}^{\ \circ}$}}\!\!\!\!\Delta}
\newcommand{\zna}{{\raise1.0ex\hbox{${}^{\ \circ}$}}\!\!\!\!\!\nabla}
\newcommand{\zS}{{\raise1.0ex\hbox{${}^{\ \circ}$}}\!\!\!\!\!S}

\newcommand{\cocal}{\textsc{cocal}\xspace}

\newcommand{\lorene}{\textsc{lorene}\xspace}

\newcommand{\sgrid}{\textsc{sgrid}\xspace}

\newcommand{\spec}{\textsc{s}p\textsc{ec}\xspace}

\newcommand{\illinois}{\textsc{Illinois GRMHD}\xspace}
\newcommand{\twopunctures}{\textsc{TwoPunctures}\xspace}

\newcommand{\GK}{\kappa}

\newcommand{\GR}{\rho}
\newcommand{\GS}{\sigma}
\newcommand{\GT}{\tau}

\newcommand{\GO}{\omega}

\newcommand{\GP}{\phi}

\newcommand{\be}{\begin{equation}}
\newcommand{\ee}{\end{equation}}

\def\QEQ{{%
    \setbox0\hbox{$I$}%
    \rlap{\hbox to \wd0{\hss--\hss}}\box0
}}

\begin{document}

\title{Great Impostors: Extremely Compact, Merging Binary Neutron Stars in the Mass Gap
Posing as Binary Black Holes}

\author{Antonios Tsokaros}
\affiliation{Department of Physics, University of Illinois at Urbana-Champaign, Urbana, IL 61801, USA}
\email{tsokaros@illinois.edu}

\author{Milton Ruiz}
\affiliation{Department of Physics, University of Illinois at Urbana-Champaign, Urbana, IL 61801, USA}
\author{Stuart L. Shapiro}
\affiliation{Department of Physics, University of Illinois at Urbana-Champaign, Urbana, IL 61801, USA}
\affiliation{Department of Astronomy \& NCSA, University of Illinois at Urbana-Champaign, Urbana, IL 61801, USA}
\author{Lunan Sun}
\affiliation{Department of Physics, University of Illinois at Urbana-Champaign, Urbana, IL 61801, USA}
\author{K\=oji Ury\=u}
\affiliation{Department of Physics, University of the Ryukyus, Senbaru, Nishihara, Okinawa 903-0213, Japan}

\date{\today}

\begin{abstract}
Can one distinguish a binary black hole undergoing a merger from a binary
neutron star if the individual compact companions have masses that fall inside
the so-called mass gap of $3-5\ M_\odot$? For neutron stars, achieving such
masses typically requires extreme compactness and in this work we present
initial data and evolutions of binary neutron stars initially in
quasiequilibrium circular orbits having a compactness $C=0.336$. These are the
most compact, nonvacuum, quasiequilibrium binary objects that have been
constructed and evolved to date, including boson stars. The compactness
achieved is only slightly smaller than the maximum possible imposed by
causality, $C_{\rm max}=0.355$, which requires the sound speed to be less than
the speed of light. By comparing the emitted gravitational
waveforms from the late inspiral to merger and postmerger phases between such
a binary neutron star vs a binary black hole of the same total mass we
identify concrete measurements that serve to distinguish them.  With that level
of compactness, the binary neutron stars exhibit no tidal disruption up until
merger, whereupon a prompt collapse is initiated even before a common core forms.
Within the accuracy of our simulations the black hole remnants from both
binaries exhibit ringdown radiation that is not distinguishable from a perturbed
Kerr spacetime.  However, their inspiral leads to phase differences of the order
of $\sim 5$ rad over an $\sim 81$ km separation (1.7 orbits) while typical neutron 
stars exhibit phase differences of $\geq 20$ rad. Although a difference of
$\sim 5$ rad can be measured by current gravitational wave laser interferometers 
(e.g., aLIGO/Virgo), uncertainties in the individual masses and spins will likely 
prevent distinguishing such compact, massive neutron stars from black holes.
\end{abstract}

\maketitle

\textit{Introduction.}\textemdash
Determining the neutron star (NS) maximum mass is a one of the most fascinating, unresolved
issues in modern astrophysics. The answer is intimately related to identifying the correct
equation of state (EOS) that describes matter at supranuclear densities \cite{Lattimer:2012nd}.
Currently the highest observed NS mass is
$2.14^{+0.20}_{-0.18}\ M_\odot$ 
\cite{Cromartie:2019kug}.
In principle, the upper limit allowing only for causality and a matching density to a well-understood
EOS somewhere around nuclear density, 
can be as high as $4.8\ M_\odot$ \cite{HARTLE1978201}, while recent studies based on the 
detection of the gravitational wave (GW) signal GW170817 place it around 
$\sim 2.2-2.3 M_\odot$ \cite{Ruiz:2017due,Shibata:2019ctb,Rezzolla:2017aly,Margalit:2017dij}.
All these studies adopt a number of underlying assumptions whose validity will require 
new observations to be verified or modified accordingly. Observationally, merging binary black 
holes (BHBHs), black hole-neutron stars (BHNSs) or binary neutron stars 
(NSNSs) whose companions have  masses that fall into the mass gap range $(3-5 M_\odot)$ are 
hard to distinguish \cite{Hannam_2013,Littenberg:2015tpa,Mandel:2015spa,Yang_2018}. 
The identification of a compact object becomes even more challenging when one
includes exotic configurations, such as quark stars, boson stars, etc., 
or alternative theories of gravity. 

The parameter that encodes how much mass a compact star can hold in a certain volume is
the compactness, defined as the dimensionless ratio $C = \frac{GM}{Rc^2}$.
Here $M$ is the Arnowitt-Desser-Misner (ADM) mass, and $R$ the areal (Schwarzschild)
radius of an isolated, nonrotating star with the same baryon mass. Our sun has 
$C=2\times 10^{-6}$, a small number indicative of its nonrelativistic nature, while 
the upper limit, $C=1/2$, is set by a Schwarzschild BH. Typical NSs 
have compactions around $\sim 0.1-0.2$ with the precise number determined by the as
yet unknown EOS. An extreme case is the incompressible fluid limit that yields 
$C=4/9=0.4\bar{4}$, the so-called Buchdahl limit \cite{Buchdahl}. This limit 
is unrealistic since it predicts an infinite sound speed. If one satisfies
the causality criterion for the sound speed (i.e. $c_s\leq c$) then the upper 
limit for compactness drops to $C_{\rm max}=0.355$
\cite{Haensel1989,Koranda1997799, Lattimer:2010uk}.

Compact binary systems provide some of the best laboratories to test the predictions of 
general relativity, 
as well as to probe possible deviations from its description of strong gravity. 
Despite the large progress that has been achieved in numerical
relativity we are still lacking  theoretical simulations that involve extremely compact 
NSs in binaries. In Ref. 
\cite{Taniguchi_2010} NSNS initial data in quasiequilibrium were constructed with 
compactness up to $C=0.26$ using the \lorene code \cite{PhysRevD.66.104019,PhysRevD.63.064029}.
Similarly, in Ref. \cite{Henriksson:2014tba} BHNS initial data were presented using 
the \spec code \cite{PhysRevD.77.124051,PFEIFFER2003253} that reach the same compactness.
Recently \cite{Tichy:2019ouu}, NSNS initial data with compactness up to $C=0.284$, together
with preliminary evolution simulations, were computed using the \sgrid code 
\cite{Tichy:2009yr,Tichy:2011gw}. 

The purpose of this work is to quantify the difference between a BHBH and an NSNS system
when the total ADM mass falls inside the mass gap and to provide useful GW diagnostics 
that may distinguish them. First, we 
construct the most massive NSNSs in quasicircular orbit with the highest compactness 
to date using our initial data solver \cocal 
\cite{Tsokaros:2015fea,Tsokaros:2018dqs,Uryu:2011ky}. The system has ADM mass $M=7.90 M_\odot$ 
and 
each star a compactness of $C=0.336$. This value (which is even higher than the maximum possible 
compactness that can be achieved by solitonic boson stars \cite{Palenzuela:2017kcg}) is only 
slightly smaller than the limiting compactness $C_{\rm max}=0.355$ set by causality. Second, using the 
\illinois code \cite{Etienne:2012te,UIUC_PAPER1, UIUC_PAPER2,prs15}, we evolve this NSNS system
and perform a detailed comparison of the gravitational waveforms with a BHBH system having the 
same initial ADM mass.
We find that an NSNS system having the above compactness inspirals very similarly to the BHBH system
and merges \textit{without essentially any tidal disruption}. 
We conjecture that to be true irrespective of the EOS for this level of compaction.
The merged NSNS remnant collapses to a BH even before a common core forms.
Since there is no disk formation, and a negligible escaping mass, one may not expect a sGRB 
or a kilonova from such an NSNS event.
The GW phase difference at the peak GW amplitude of the NSNS system is
$\sim 5$ 
rad with respect to the  BHBH binary inside the 
band $[0.6,1]\ {\rm KHz}$. This phase difference corresponds to $\sim 20\%$ 
of the accumulated phase 
during the last $\sim 1.7$ orbits (corresponding to an initial separation of $\sim 81$ km) 
and \textit{can} be detected by the aLIGO/Virgo network. On the other hand the postmerger 
remnants have ringdown waveforms that \textit{cannot} be distinguished from the Kerr BH ringdown 
within the accuracy of our  simulations.

In the following we employ geometric units in which $G=c=M_\odot=1$, unless stated otherwise.

\textit{EOS and numerical methods.}\textemdash
In this work we employ the cold EOS adopted in Ref. 
\cite{Tsokaros:2019mlz} which we called ALF2cc. It is based on the ALF2 EOS 
\cite{Alford2005} where the region with rest-mass density $\GR_0\geq\GR_{0s}$ is
replaced by the maximum stiffness EOS given by
\be
P=\GS(\GR-\GR_s) + P_s \, .
\label{eq:eoscc}
\ee
Here $\GS$ is a dimensionless parameter, $\GR$ is the total energy density,
and $P_s$ the pressure at $\GR_s$.
The solutions presented in this work assume $\GS=1.0$,
i.e. a core at the causal limit, which represents the maximally compact,
compressible EOS \cite{2016PhR...621..127L}. The matching density $\GR_s$
is, in principle, the point beyond which current nuclear studies cannot confidently 
describe matter and it is a multiple of nuclear matter density 
$\GR_{0\rm nuc}=2.7\times 10^{14}\ {\rm gr/cm^3}$.
In our current study though, we simply take $\GR_{0s}=\GR_{0\rm nuc}$ in
order to maximize the NS compactness and thereby provide a benchmark
upon which future studies can be compared (see also Refs. \cite{2009MNRAS.398L..31L, Zhou:2017xhf} 
for other EOSs that support such high compactions).

\begin{figure}                                                                   
\begin{center}                                                                   
\includegraphics[width=0.99\columnwidth]{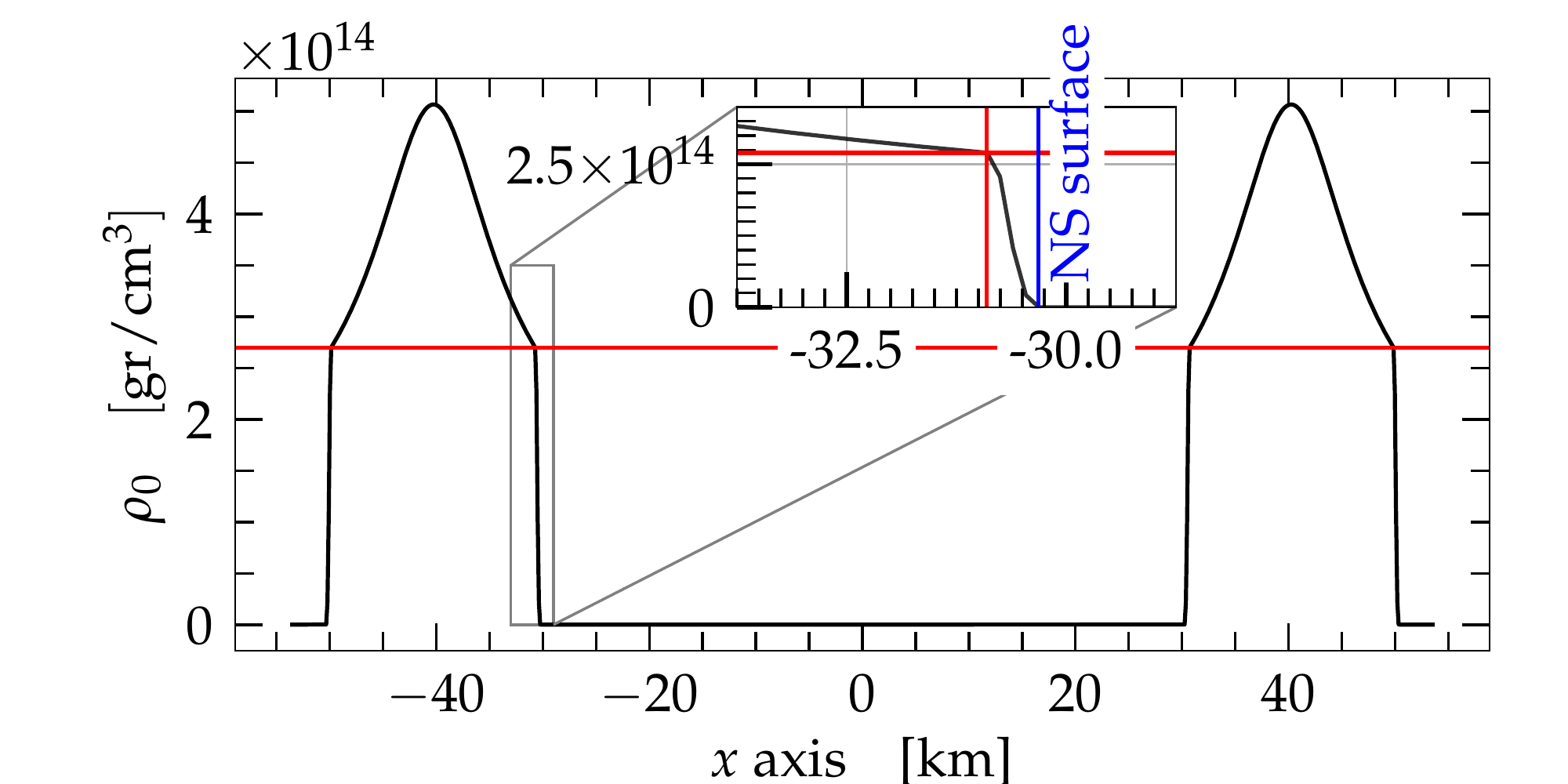}                            
\caption{Rest-mass density profile across the $x$ axis for the NSNS system with the
ALF2cc EOS. Horizontal red line corresponds to nuclear density
$\GR_{0\rm nuc}$. The inset enlarges the area close to the surface where
the density drops from $\GR_{0\rm nuc}$ to zero in a steep manner.}  
\label{fig:rho0}                                                                 
\end{center}                                                                     
\end{figure}

Our NSNS initial data are computed using the \cocal
code \cite{Tsokaros:2015fea, Tsokaros:2016eik, Tsokaros:2018dqs,Uryu:2011ky}
while the BHBH initial data using the \twopunctures code \cite{abt04,Paschalidis:2013oya}.
For the NSNS binary each NS has a 
rest mass $M_0=5.18$ which corresponds to a spherical star with
compactness $M/R=0.336$. The ADM mass of the system is $M=7.90$, the
coordinate separation is $80.6$ km, and the orbital angular velocity
is $\Omega M=0.0460$. All NS radii are approximately 10 km.
Although the distance between the two
NSs is very large compared with typical NSNS simulations it only results
in $N\sim 1.7$ orbits prior to merger according to the lowest order post-Newtonian
(PN) formula \cite{Blanchet:2013haa},
$2\pi N=(M\Omega)^{-5/3}/(32\nu)$ [$\nu=m_1 m_2/(m_1+m_2)^2=1/4$], due to
the very large gravitational mass of the system. The challenge both for 
the initial data calculation, as well as for the evolution, is to resolve
a relatively small and very compact NS over  such a large binary separation.
In Fig. \ref{fig:rho0} we plot the rest-mass density profile across the
$x$ axis that passes through the center of each star. 
The profiles look very similar to self-bound quark stars whose
density at the surface is finite. The red horizontal line divides the
region with the causal EOS Eq. (\ref{eq:eoscc}) (above the red line) from
the polytropic ALF2 crust (below the red line). The region near the surface 
is expanded in the inset of Fig. \ref{fig:rho0} which shows that our stars
exhibit  $\sim 500$ m of crust, which is $\sim 5\%$ of their
radius but only $\leq 1\%$ of the rest mass. 
The blue vertical line pinpoints the surface of the NS while
the red lines mark the change of the EOS.

\begin{figure*}                                                                   
\begin{center}                                                                   
\includegraphics[width=0.65\columnwidth]{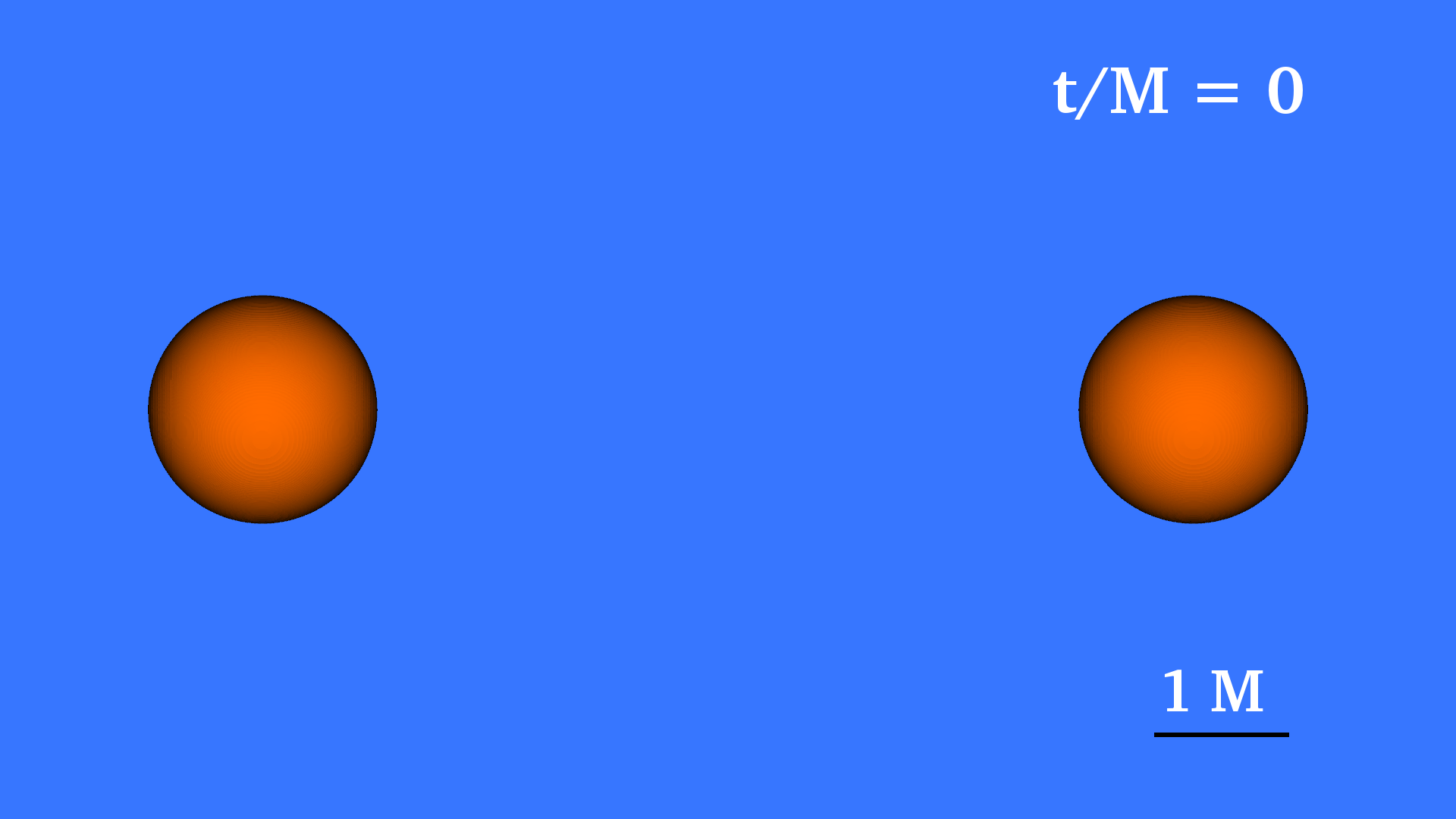}                            
\includegraphics[width=0.65\columnwidth]{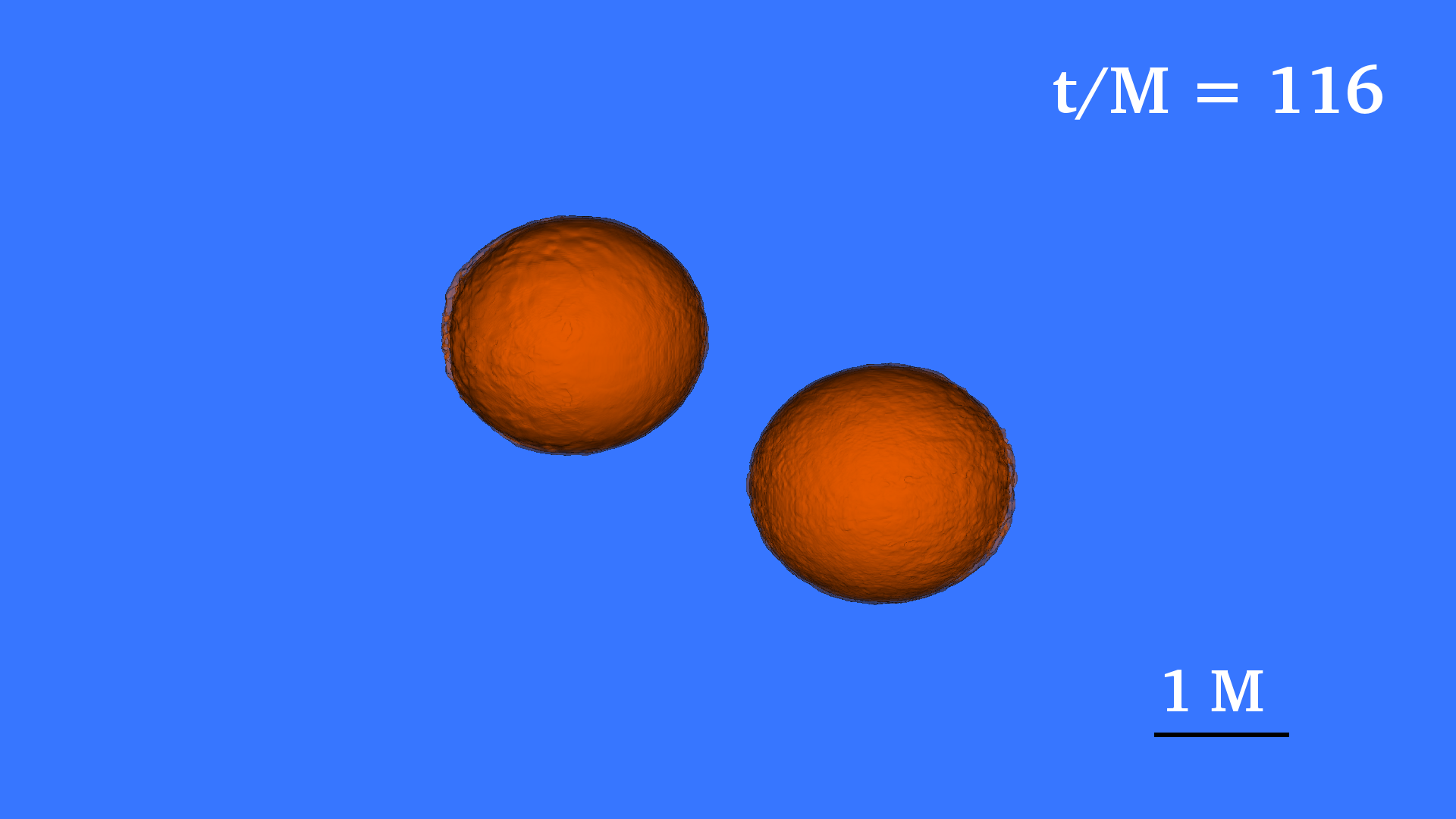}                            
\includegraphics[width=0.65\columnwidth]{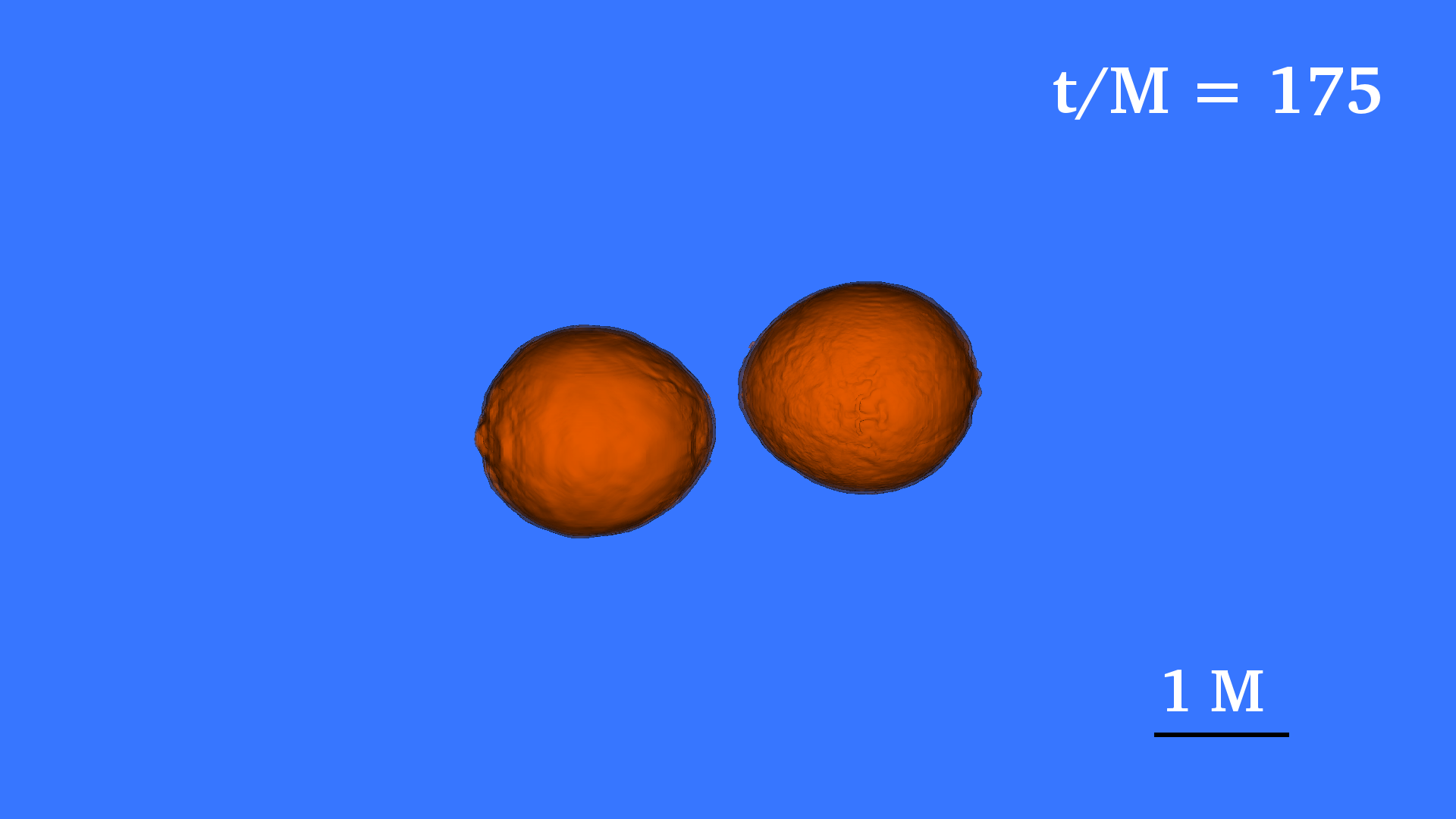}                            
\includegraphics[width=0.65\columnwidth]{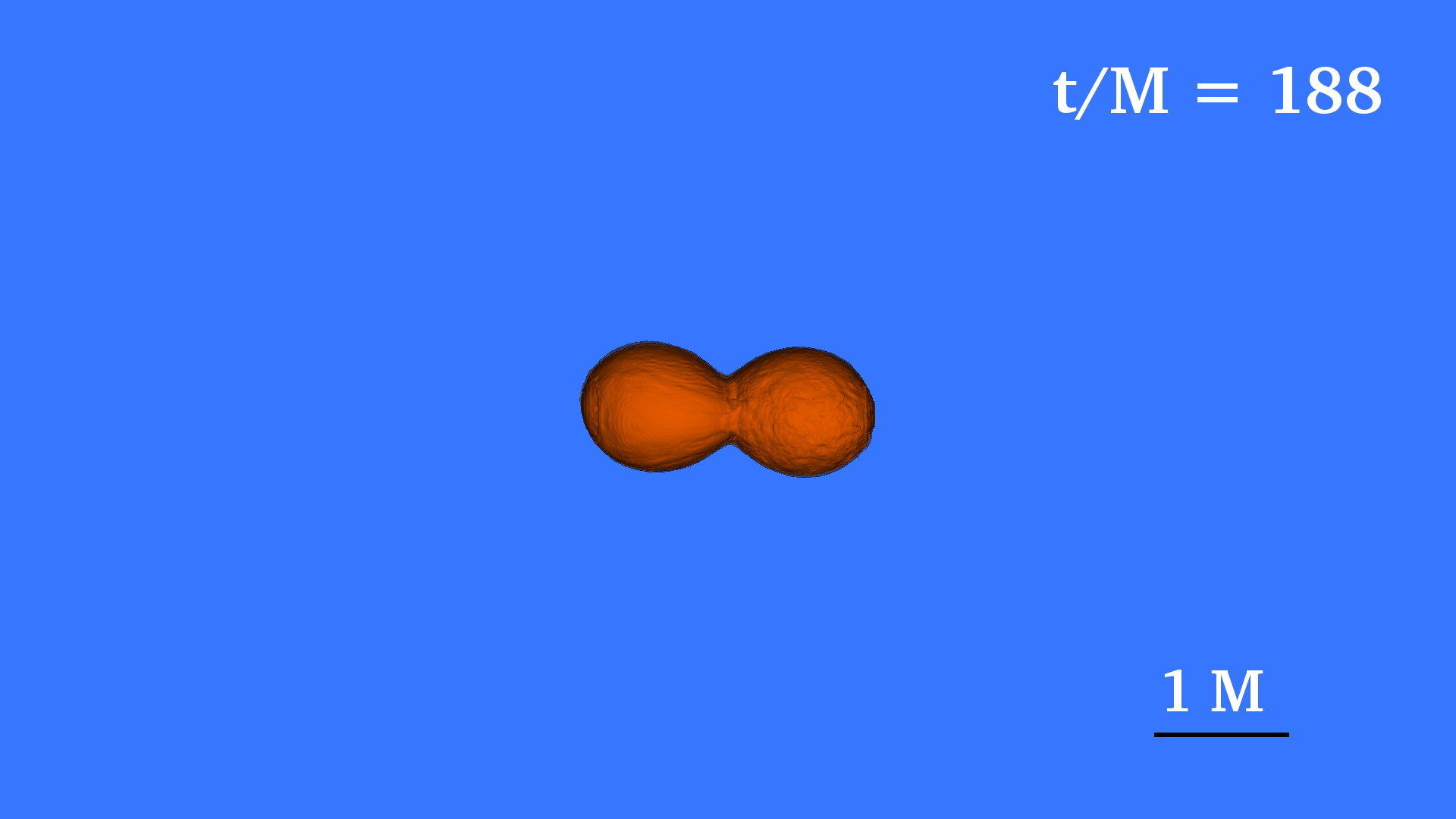}                            
\includegraphics[width=0.65\columnwidth]{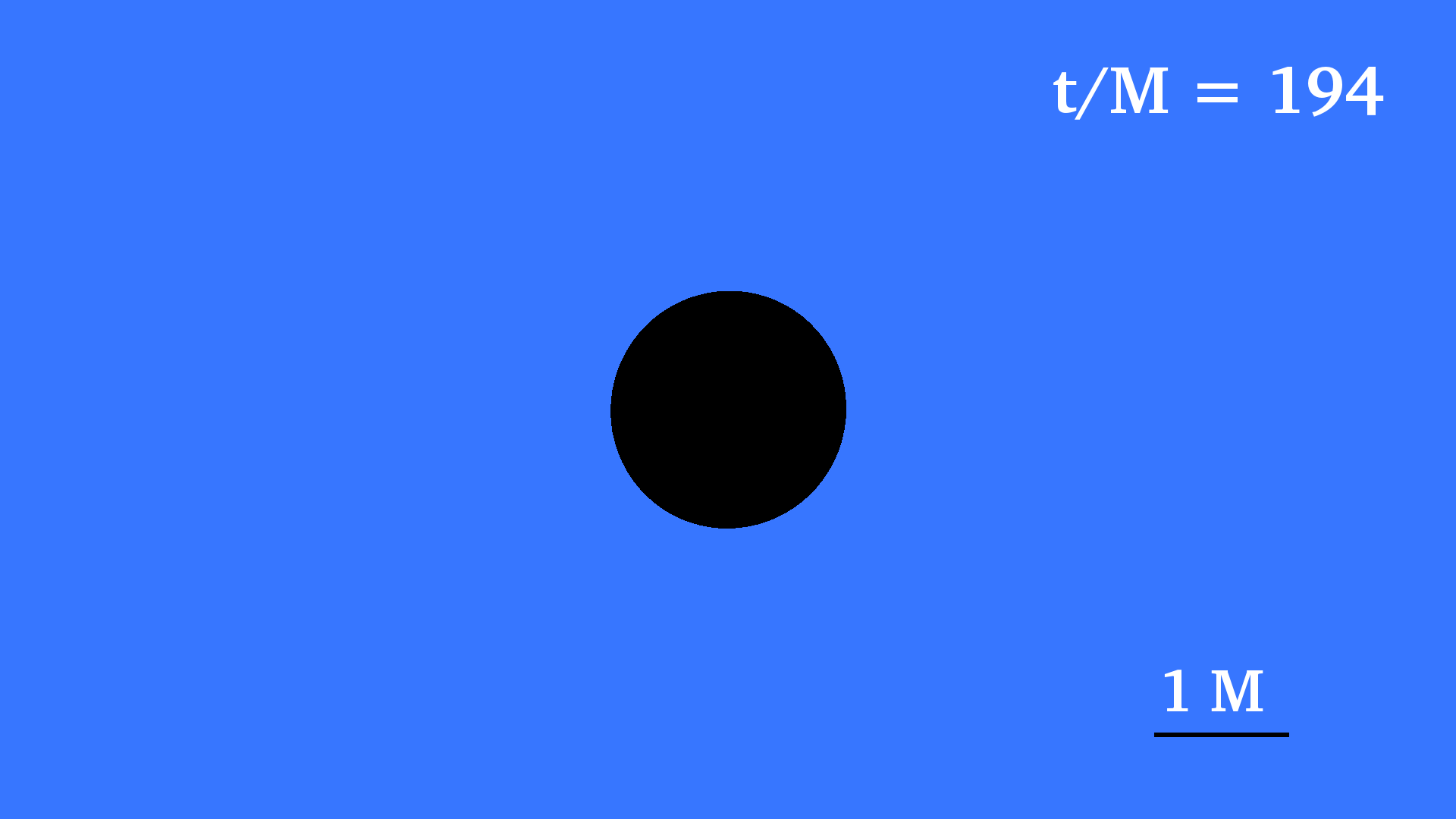}                            
\includegraphics[width=0.65\columnwidth]{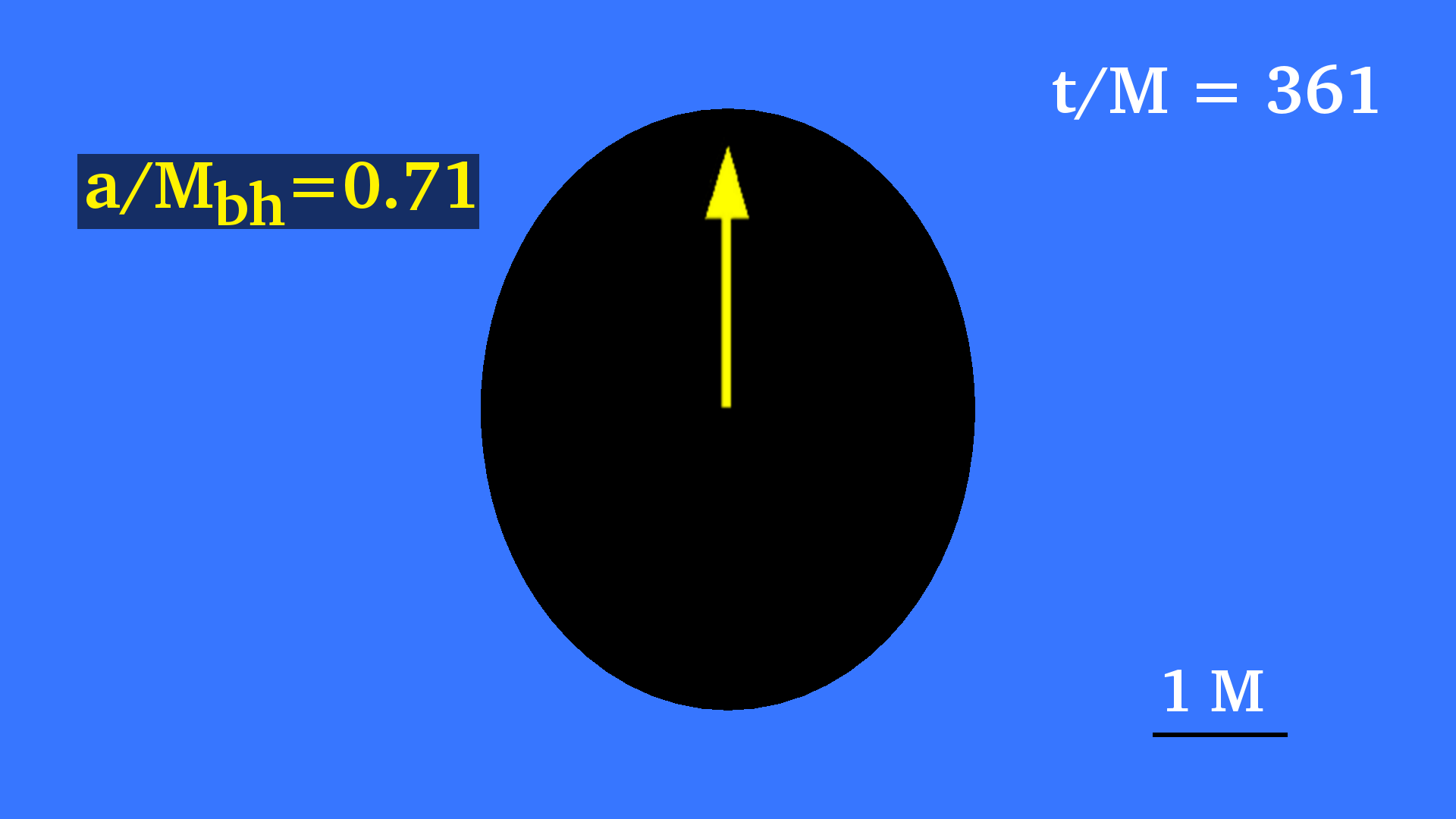}                            
\caption{ Evolution of an $M=7.90 M_\odot$ binary NSNS system.
Isocontours of rest-mass density $\GR_0=2.263\times 10^{13}\ {\rm gr/cm^3}$ 
which corresponds to  $0.998 R_x(0)$ and therefore
is a accurate representation of the surface of the star.}  
\label{fig:insp}                                                                 
\end{center}                                                                     
\end{figure*}

We perform evolutions of the BHBH and NSNS systems using the 
\illinois adaptive-mesh-refinement code \cite{Etienne:2012te,UIUC_PAPER1, UIUC_PAPER2,prs15} 
which employs the Baumgarte-Shapiro-Shibata-Nakamura (BSSN) formulation of 
Einstein's equations \cite{shibnak95,BS, BSBook} to evolve the spacetime metric
and the matter fields (see Ref. \cite{Tsokaros:2019mlz} for an evolution that also adopts the ALF2cc EOS).  
For the NSNS binary we use two resolutions: Resolution R1nsns
uses eight refinement boxes with $\Delta x_{\rm min}=118$ m, while resolution
R2nsns employs $\Delta x_{\rm min}=98.5$ m. Both R1nsns and R2nsns can resolve the crust 
by four or five points (initially). In the Supplemental Material we plot the violations of the
constraint equations, where their peak values indeed come from the crust. Future simulations will 
improve the accuracy. For the BHBH binary, resolution Rbhbh
uses nine refinement boxes with 
$\Delta x_{\rm min}=175$ m. Reflection symmetry is imposed across the orbital plane.
Both resolutions that we use are among the highest in NSNS simulations.
According to Ref. \cite{Kiuchi:2017pte} one needs $\Delta x_{\rm min} \le 100$
m to achieve sub-radian accuracy ($\sim 0.2$ rad) and nearly
convergent waveforms in approximately 15 orbits. In our case the 
high compactness of our NSs necessitate the use of such resolution,
while lower resolutions seem inadequate to keep the stars in bound orbits.

For the GW diagnostics we use the methods described in Ref. \cite{Tsokaros:2019anx}
and denote by 
$h^{\ell m}(t) = h_{+}^{\ell m}(t) - i h_{\times}^{\ell m}(t) = A_{\ell m}(t) e^{-i\Phi_{\ell m} (t)}$
the strain of the $(\ell,m)$ mode and 
$\GO_{\ell m} = 2\pi f_{\ell m} = \frac{d\Phi_{\ell m}}{dt}$
the corresponding GW frequency.

\textit{Results.}\textemdash
The evolution of the NSNS system is depicted in Fig. \ref{fig:insp} where isocontours
at density $\GR_0=2.26\times 10^{13}\ {\rm gr/cm^3} = \GR_0^{\rm max}(0)/22.4$ are
plotted at various times during the inspiral. This isocontour corresponds to the
density at a radial distance $0.998 R_x(0)$ measured from the maximum density point in the NSs;
therefore it is an accurate representation of the surface 
of the star. In accordance with the (PN) prediction the binary performs approximately 
$\sim 1.7$ orbits with the two stars starting as two spherical configurations to high 
accuracy. 
Tidal distortion becomes evident only at $1.5$ orbits when $t/M\approx 170$. 
Shortly afterwards (less than a quarter of an orbit) merger begins \textit{with no cusp formation} 
(cf. Ref. \cite{Baiotti:2008ra}). The surface remains intact up until the 
merging of the two NSs at $t/M=180$ where they actually touch. Immediately thereafter the remnant 
collapses, at $t/M\sim 183$, when the structure still has a clear dumbbell shape 
as in the snapshot at the left column, bottom panel of Fig. \ref{fig:insp} (with $t/M=188$).
The apparent horizon immediately after collapse has a spherical shape but settles as a prolate 
configuration at the end of the simulation. This is simply a gauge effect caused by the NSNS
moving puncture coordinates. The ratio of polar to equatorial proper circumferences asymptotes to 
$\sim 0.89<1$ (see postmerger section) \cite{movies}.

\begin{figure} 
\begin{center}                                                                   
\includegraphics[width=0.99\columnwidth]{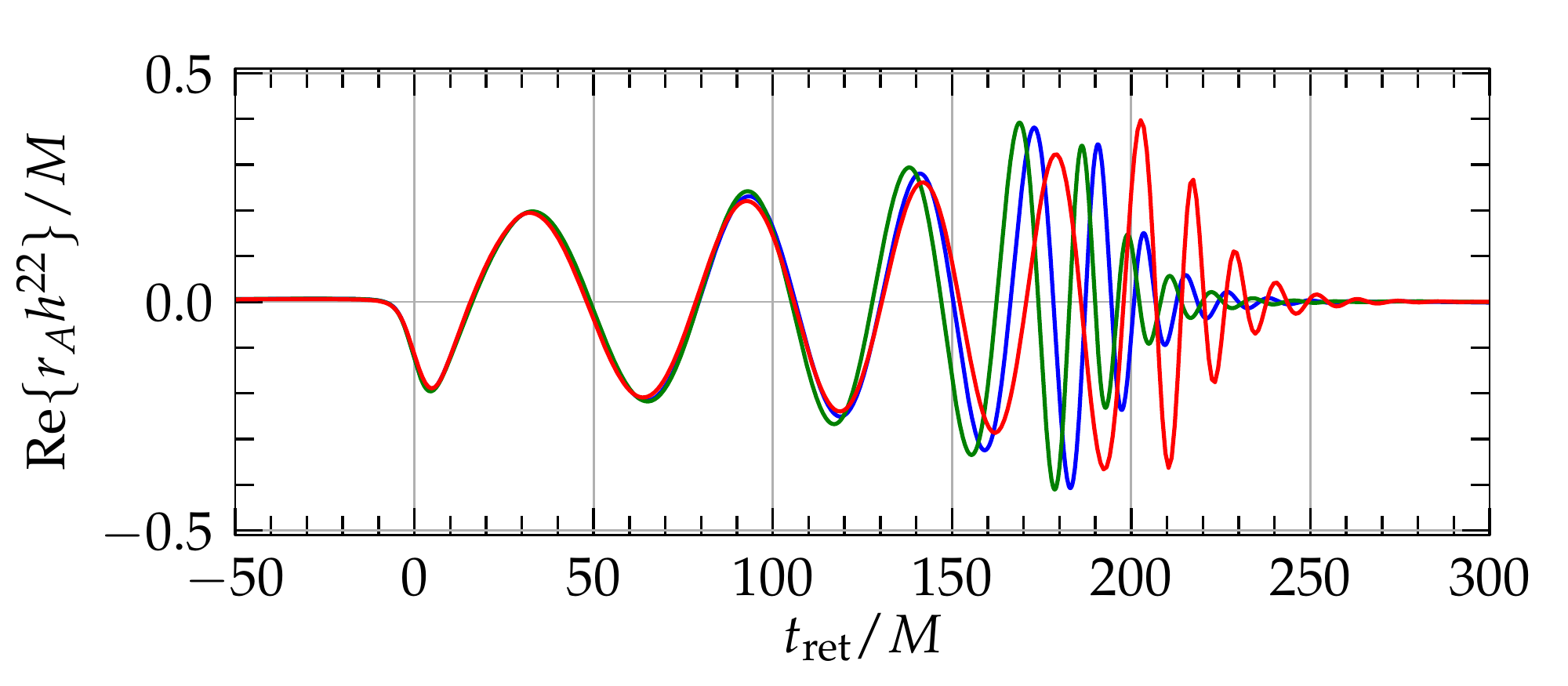}                            
\includegraphics[width=0.99\columnwidth]{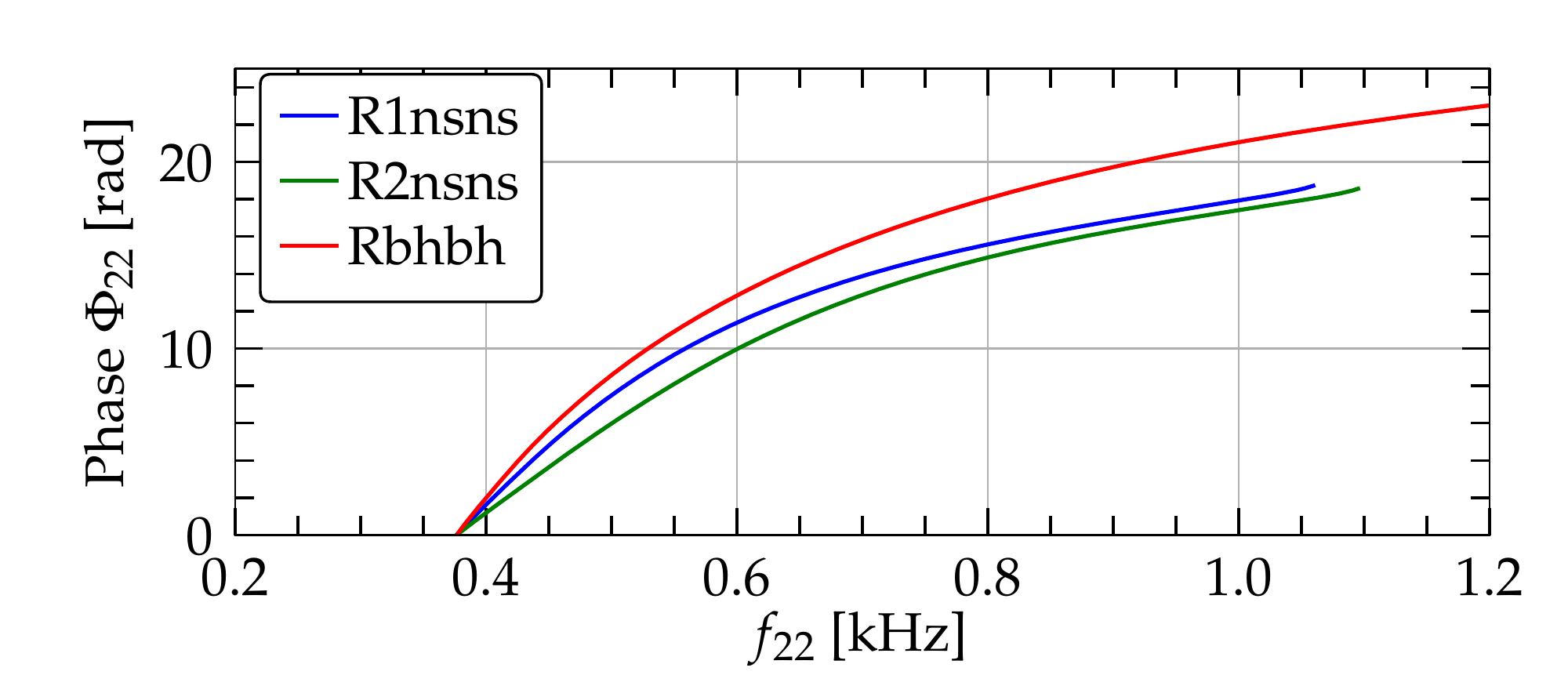}                            
\caption{Top panel: strain vs retarded time for the $(\ell=2,m=2)$ dominant mode. 
R1nsns and R2nsns correspond to the two resolutions of the NSNS system 
and Rbhbh to the BHBH simulation. Bottom panel shows the phase $\Phi_{22}(t)$ of the strain up until
its maximum vs the GW frequency.    }  
\label{fig:gw}                                                                 
\end{center}                                                                     
\end{figure}

\textit{Inspiral.}\textemdash
In the top panel of Fig. \ref{fig:gw} we plot the normalized strain of the (2,2) mode 
($r_A$ is the areal radius).
Despite the small number of orbits performed by our NSNS system one can appreciate the
fact that the early part of the inspiral is very similar for both NSNS and BHBH, with differences
starting to appear when the tidal interactions begin at $t/M\sim 130$. This
observation is expected since our NSs are very compact and thus have small radii $R_x$;
at initial separations of  $\sim 8R_x$ the inspiral should develop according to the point-particle approximation.
For the same reason one expects that tidal effects will start to develop later on, 
which is what we find. For all simulations we extract $\Psi_4$ at seven different 
radii from the orbital center and verify the expected ``peeling'' property $r\Psi_4=$const. In the figures shown here 
we used an extraction radius $R_{\rm ex}=106 M = 1241$ km. 

As expected the NSNS binary merges earlier than the BHBH binary even though tidal effects are minimal.
Also, by comparing the two NSNS resolutions, we observe that the NSNS evolved with the high 
resolution (R2nsns) merges slightly earlier than R1nsns. In the bottom panel of Fig. \ref{fig:gw}
we plot the phase of the GW strain $\Phi_{22}$ of the (2,2) mode as a function of frequency
up until the moment of maximum strain. At a given time the angular velocity of the BHBH binary is 
smaller than the corresponding of the NSNS binary, leading to delayed merger. Although our two NSNS 
resolutions produce a small phase difference between them, the important quantity here is the phase 
difference $\Delta\Phi_{22}=|\Phi_{22}^{^{\rm BHBH}}-\Phi_{22}^{^{\rm NSNS}}|$. 
The largest dephasing between the BHBH and NSNS curves is $\leq 3.5$ rad within $[0.6,1.0]$ KHz. 
If we extrapolate our results to infinite resolution 
(see Refs. \cite{Etienne:2007jg,els10,Tsokaros:2019anx,Espino:2019xcl} and the Supplemental Material)
we conclude that a BHBH binary will have maximum  $\sim 5$ 
rad difference with respect to an NSNS 
of compaction $C=0.336$ in the aforementioned bandwidth. 
This result is in accordance with other studies \cite{Dietrich:2019kaq,Tsokaros:2019anx}
where typical NSNS binaries are employed and phase 
differences $\gtrapprox 20$ rad were recorded depending on the EOS, and the NSNS binary properties.
Our maximally compact stars yield a minimal but \textit{measurable} phase difference.
Given the fact that this phase difference is produced in the last $\sim 1.7$ orbits, or through an 
accumulated phase of only $\sim 20$ rad, we calculate that the dephasing relative to the BHBH case 
is significant $\sim 20\%$.

\begin{figure}                                                                   
\begin{center}                                                                   
\includegraphics[width=0.99\columnwidth]{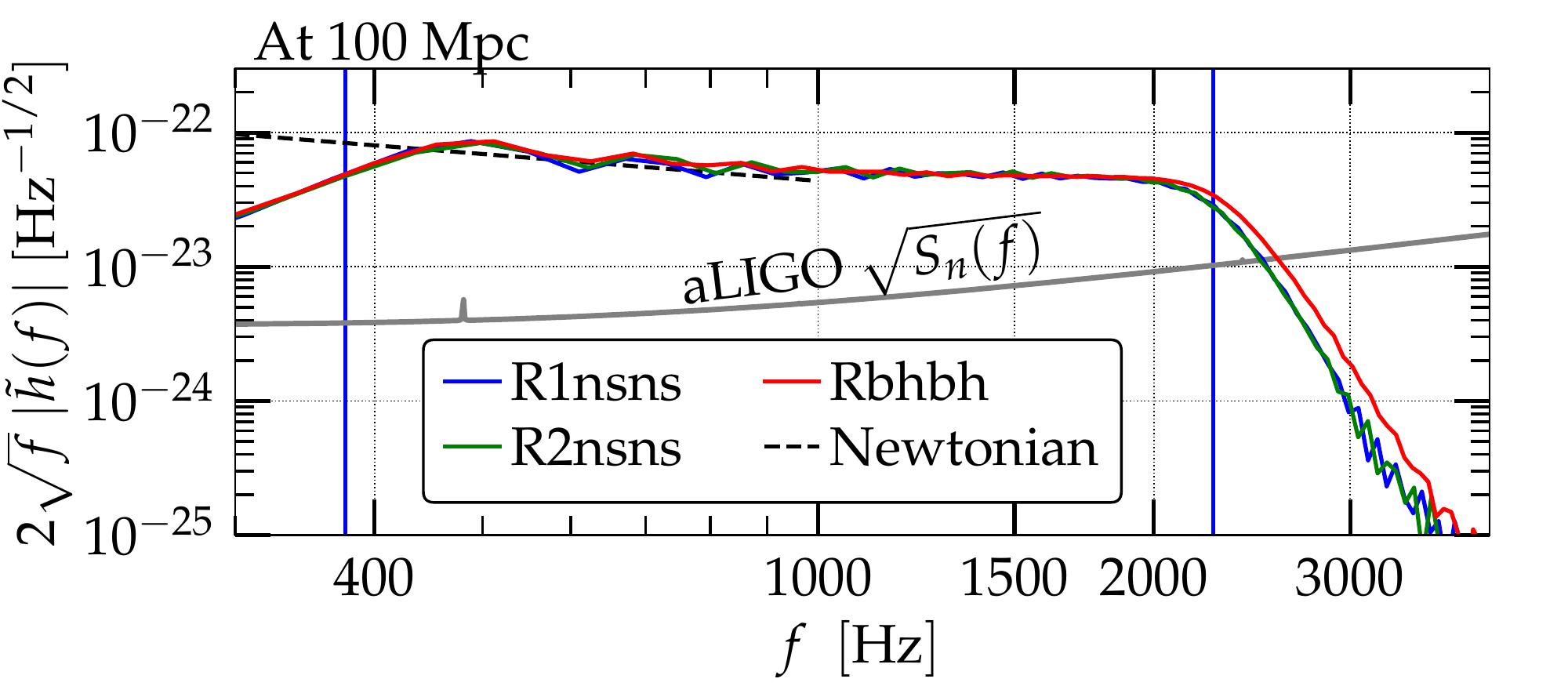}                            
\includegraphics[width=0.99\columnwidth]{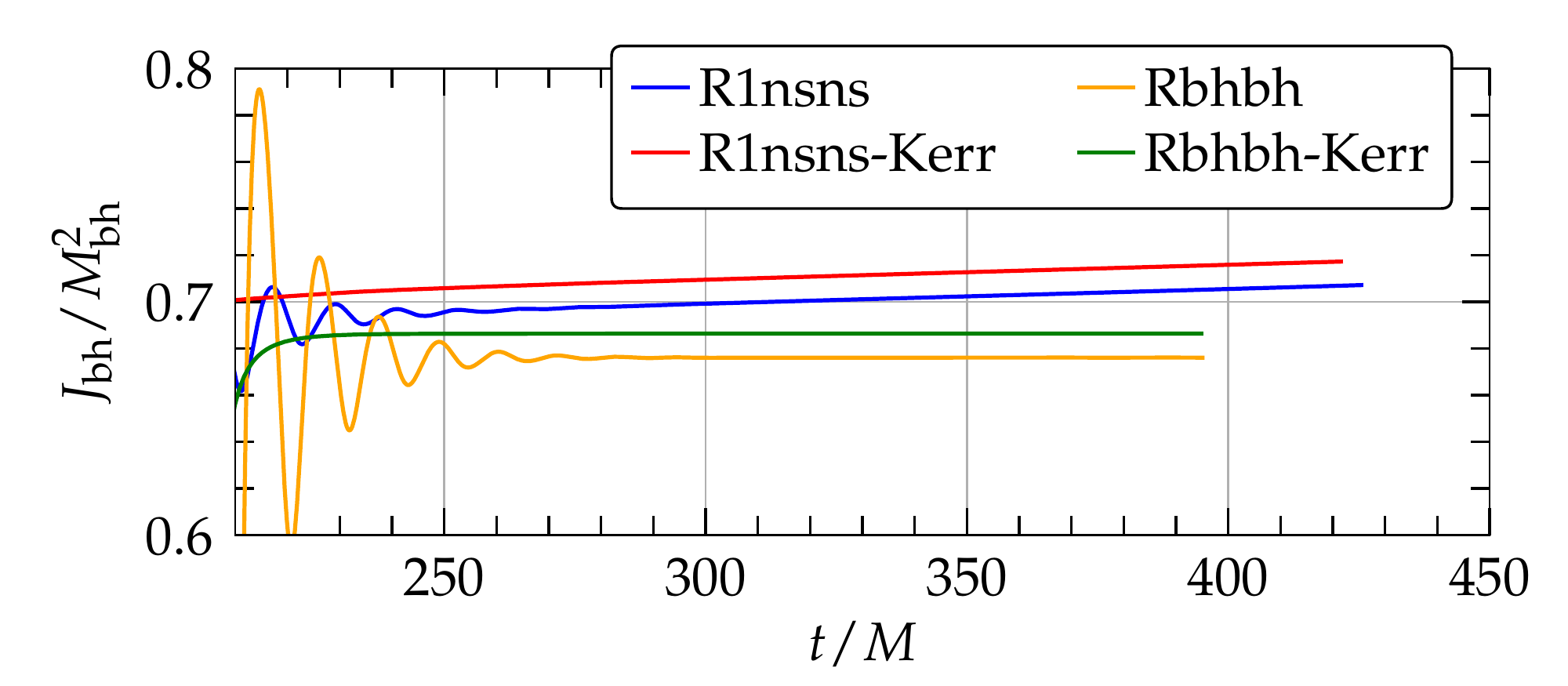}                            
\caption{ Top panel: Fourier spectra of numerical waveforms (colored lines).
Vertical blue lines correspond to the frequency of the (2,2) mode at the initial separation 
and at the ringdown. Bottom panel: Spin of the remnant BHs.}
\label{fig:fft}                             
\end{center}                                                                     
\end{figure}

In Fig. \ref{fig:fft} top panel we plot the Fourier spectrum at 100 Mpc of the (2,2) mode 
$\tilde{h}(f)=\mathcal{F}(h^{22})$
for the NSNS (two resolutions, blue and green lines) and BHBH (red line) binaries. 
We also plot the aLIGO noise curve (gray line) \cite{zero_detuning}, as well as the
Newtonian prediction \cite{Cutler:1994ys} (dashed lines).
Vertical blue lines correspond to the initial and ringdown GW frequency
of the (2,2) mode (see next section). The power spectral density for the two kind of
binaries is very similar, thus in order to quantify their difference we compute
the match function \cite{Allen2012}
$\mathcal{M} =  \underset{(\GP_c,t_c)}{{\rm max}}
\frac{(\tilde{h}_1|\tilde{h}_2(\GP_c,t_c))}
     {\sqrt{(\tilde{h}_1|\tilde{h}_1)(\tilde{h}_2|\tilde{h}_2)}}$
where the maximization is taken over a large set of phase shifts $\GP_c$ and time shifts $t_c$.
Here $(\tilde{h}_1|\tilde{h}_2)$ denotes the standard noise-weighted inner product \cite{Allen2012}.
For both resolutions R1nsns and R2nsns we find $\mathcal{M}=0.998$, i.e. the waveforms 
emitted at $100$ Mpc are
distinguishable with current detectors for a signal-to-noise ratio of, e.g., 25 
\cite{Harry_2018,PhysRevD.95.104004}, comparable to GW150914 \cite{firstBHBH}.
However, uncertainties in the 
individual masses and spins 
will likely prevent these detectors from
distinguishing these compact, massive NS binaries from BH binaries.

%

\textit{Postmerger.}\textemdash
In order to diagnose the spin of the remnant BHs we use two methods. First, using the
isolated horizon formalism \cite{dkss03}
we calculate 
$a/M_{\rm bh} = J_{\rm bh}/M_{\rm bh}^2$ (using R1nsns and Rbhbh for the NSNS and BHBH runs, 
respectively). Second, for the Kerr spacetime the ratio of 
proper polar horizon circumference, $L_p$, to the equatorial one, $L_e$ is
$\frac{L_p}{L_e} = 4\sqrt{r_{+}^2+a^2}\ E\left(\frac{a^2}{r_{+}^2+a^2}\right)$,
where $E(x)$ is the complete elliptic integral of the second kind.
Following Ref. \cite{Brandt1995} we can approximate this expression by
$L_p/L_e \approx [\sqrt{1-(a/M_{\rm bh})^2}+1.55]/2.55$. By computing the ratio $L_p/L_e$
directly from the metric one can get an estimate of $a/M_{\rm bh}$ using the latter approximate formula. This
quantity is plotted in the bottom panel of Fig. \ref{fig:fft}.
The two spin diagnostics agree to a level of $\sim 1.5\%$ for both binaries,
while the spin of the BHBH binary remnant differs from the one of the NSNS binary
remnant by $\sim 4\%$ with the NSNS remnant BH having higher spin. This is 
consistent with the calculated GW angular momentum emission and the conservation of 
angular momentum diagnostic shown in the Supplemental Material.

Although the mode frequencies $\GO_{\ell m}$ 
of the GW signal during inspiral and merger are described by complicated functions of time, 
during ringdown the GW signal can be described with high accuracy as a simple superposition 
of damped sinusoids characterized by three indices: the two spherical harmonic indices $\ell,\ m$ 
and a third overtone index, $n=0,1,\ldots$, which here we assume to be the fundamental one
$n=0$ \cite{Kokkotas:1999bd,Berti:2005ys}. As a consequence of the no-hair theorem, all 
dimensionless mode frequencies, $M_{\rm bh}\GO_{\ell mn}$, and damping times 
$\GT_{\ell mn}/M_{\rm bh}$ depend only on the dimensionless spin of the remnant BH. 
For our BHBH and NSNS binaries the dominant modes are the (2,2) and (4,4) ones. 
The frequencies of the (2,2) modes for the NSNS and BHBH binaries are very 
close to each other and lead to a BH spin which is consistent
with the values presented in Fig. \ref{fig:fft} and discussed in the previous paragraph.
In particular we find that $(M_{\rm bh}\GO_{22})_{_{\rm NSNS}}=0.54$ while 
$(M_{\rm bh}\GO_{22})_{_{\rm BHBH}}=0.53$. The (4,4) modes are more noisy and yield
$(M_{\rm bh}\GO_{44})_{_{\rm NSNS}}=1.13$ and $(M_{\rm bh}\GO_{44})_{_{\rm BHBH}}=1.11$. 

In order to compare the ringdown of our models to well-known results from perturbation
theory for Kerr BHs we use the fits provided in Ref. \cite{Berti:2005ys}:
\begin{eqnarray}
M_{\rm bh}\omega_{22} & = & 1.5251 - 1.1568(1-a/M_{\rm bh})^{0.1292}   \\
M_{\rm bh}\omega_{44} & = & 2.3000 - 1.5056(1-a/M_{\rm bh})^{0.2244}  \nonumber \ . 
\end{eqnarray}
Using the (2,2) frequencies
of our models we find $(a/M_{\rm bh})_{_{\rm NSNS}}=0.71$ and $(a/M_{\rm bh})_{_{\rm BHBH}}=0.69$
in very good agreement with the values shown in Fig. \ref{fig:fft}. The (4,4) mode 
predicts spins which differ by $\sim 5\%$ from the ones coming 
from the (2,2) one. This is within the accuracy of our angular momentum conservation 
(see Supplemental Material) but is also expected due to the larger noise in our simulation for those
higher order modes. In conclusion, the ringdown of both the NSNS and BHBH binaries is
consistent with the ringdown of a perturbed Kerr BH, with the mass and angular momentum of
the remnants  closely matching the ones predicted by the Kerr metric.

\vspace{1cm}

It is a pleasure to thank J. Creighton and N. Yunes for useful discussions.
We thank the Illinois Relativity Group REU team, G. Liu, K. Nelli, and  M. N.T
Nguyen for assistance in creating Fig. \ref{fig:insp}.
This work was supported by NSF Grant No. PHY-1662211 and NASA Grant No. 80NSSC17K0070 to the
University of Illinois at Urbana-Champaign, as well as by JSPS Grant-in-Aid for Scientific
Research (C) 15K05085 and 18K03624 to the University of Ryukyus. This work made use of the
Extreme Science and Engineering Discovery Environment (XSEDE), which is supported by National
Science Foundation Grant No. TG-MCA99S008. This research is part of the Blue Waters
sustained-petascale computing project, which is supported by the National Science Foundation
(Grants No. OCI-0725070 and No. ACI-1238993) and the State of Illinois. Blue Waters
is a joint effort of the University of Illinois at Urbana-Champaign and its National Center
for Supercomputing Applications. Resources supporting this work were also provided by the
NASA High-End Computing (HEC) Program through the NASA Advanced  Supercomputing  (NAS)
Division at Ames Research Center.

\bibliographystyle{apsrev4-1}
\bibliography{references}

\section{Supplemental Material}

\begin{figure}                                                                   
\begin{center}                                                                   
\includegraphics[width=0.99\columnwidth]{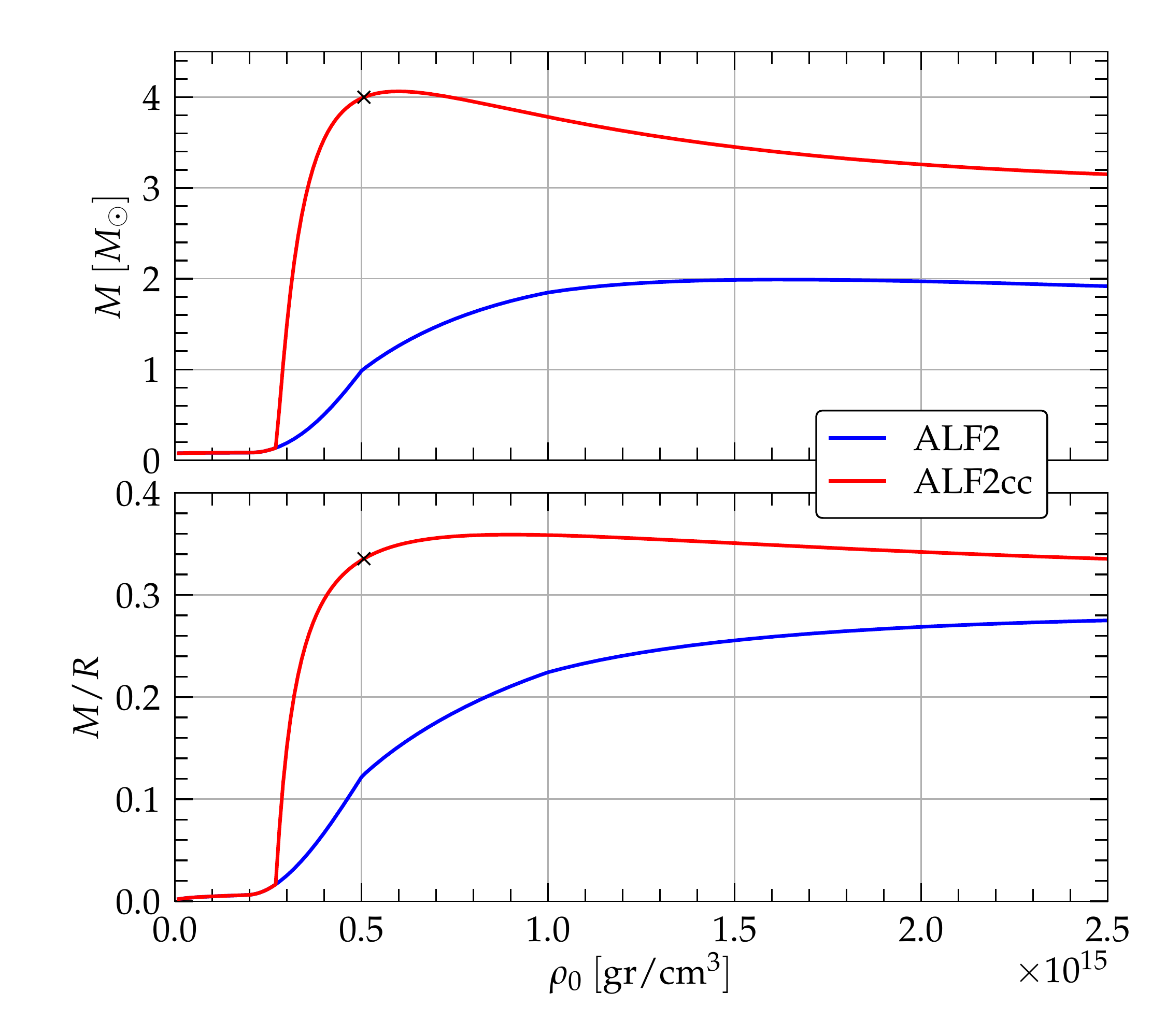}                            
\caption{Mass and compactness vs rest-mass density for the ALF2 and ALF2cc
EoSs for equilibrium models in spherical symmetry. The $\times$ marks the configuration
we choose for each compact binary neutron star companion.}  
\label{fig:ALF2cc_TOV}                                                                 
\end{center}                                                                     
\end{figure} 

\begin{figure}                                                                   
\begin{center}                                                                   
\includegraphics[width=0.99\columnwidth]{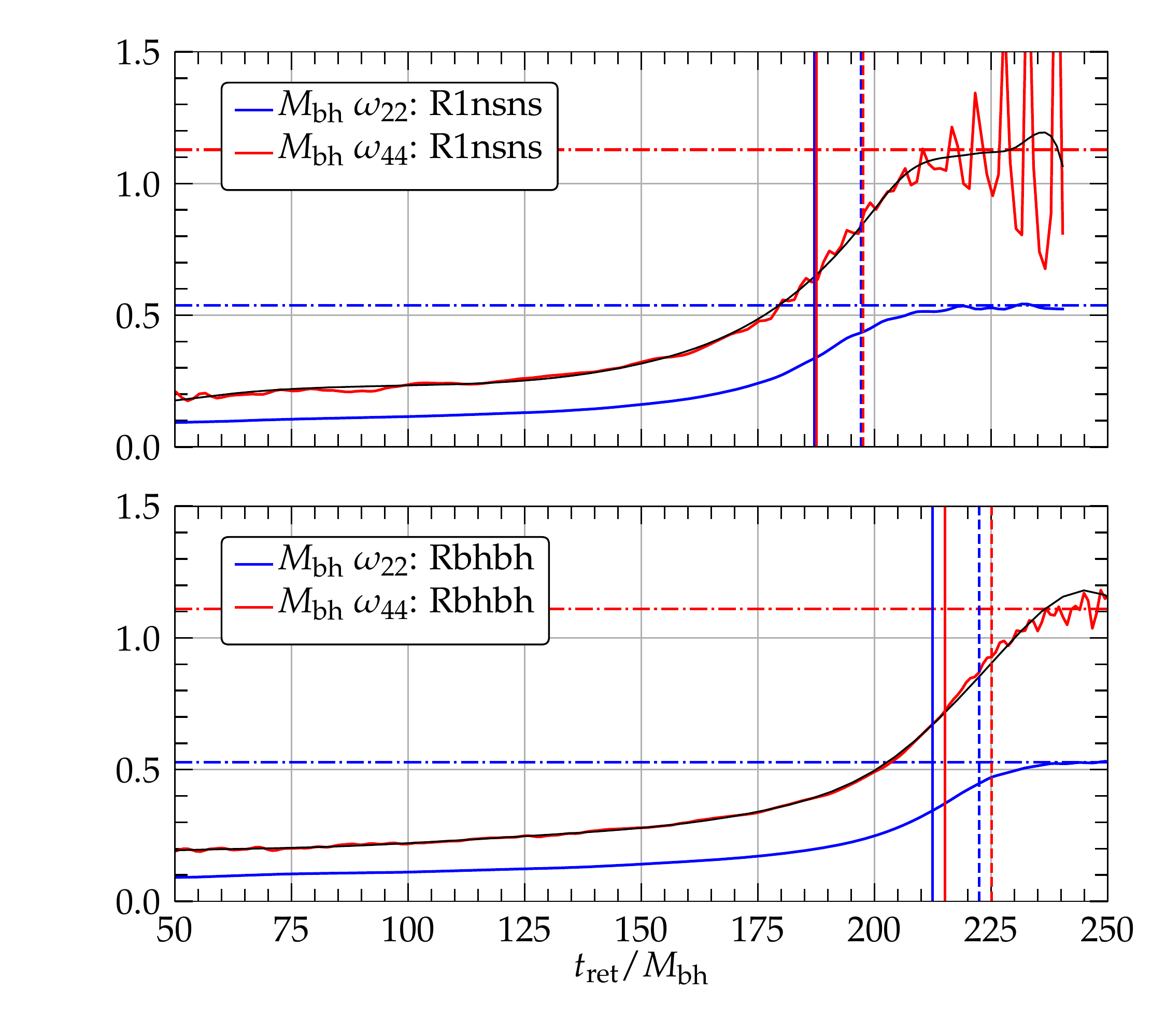}                            
\caption{Evolution of the dimensionless frequencies $M_{\rm bh}\GO_{22}$ and $M_{\rm bh}\GO_{44}$
for the two dominant modes. Top panel refers to the NSNS merger while bottom panel to the BHBH
merger. Vertical solid lines denote the time of the maximum strain for the corresponding modes (blue for
the (2,2) and red for the (4,4) mode) while vertical dashed lines correspond to the start of the ringdown,
$\Delta t_{\rm ret}=10M_{\rm bh}$ after the peak strain times. Black lines correspond to a fit while
horizontal lines to average asymptotic values. }
\label{fig:ringdown}                             
\end{center}                                                                     
\end{figure} 

\section{EoS}

\begin{figure*}                                                                   
\begin{center}                                                                   
\includegraphics[width=0.65\columnwidth]{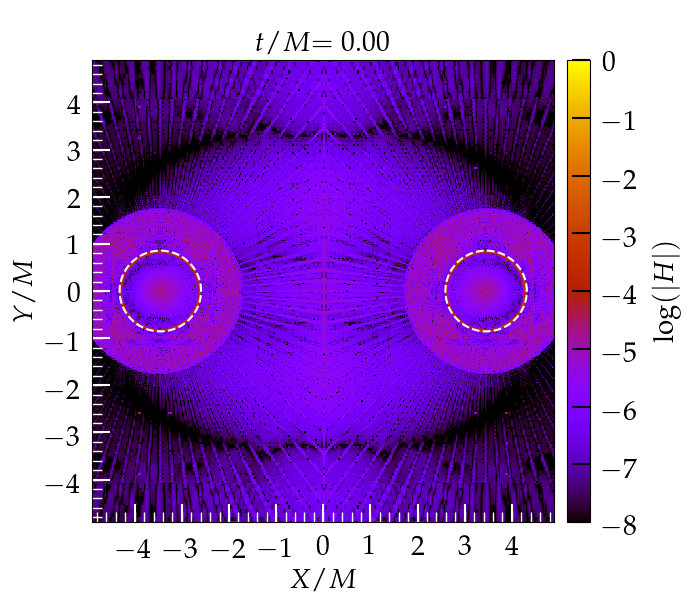}                            
\includegraphics[width=0.65\columnwidth]{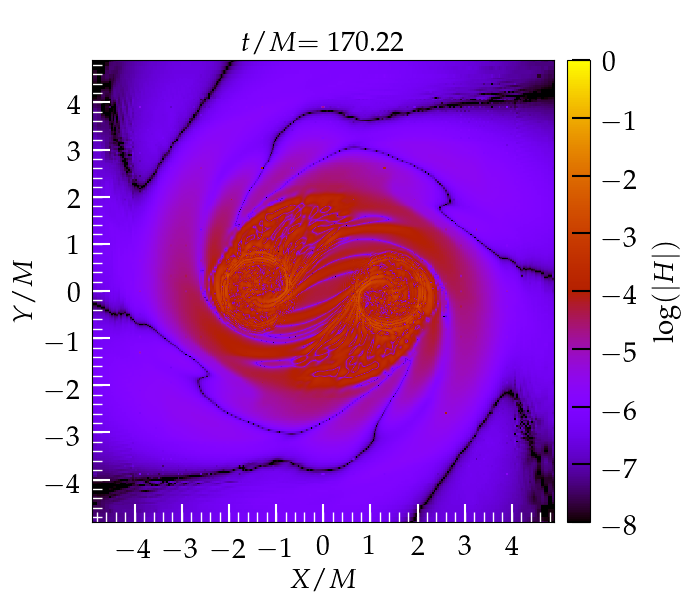}                            
\includegraphics[width=0.65\columnwidth]{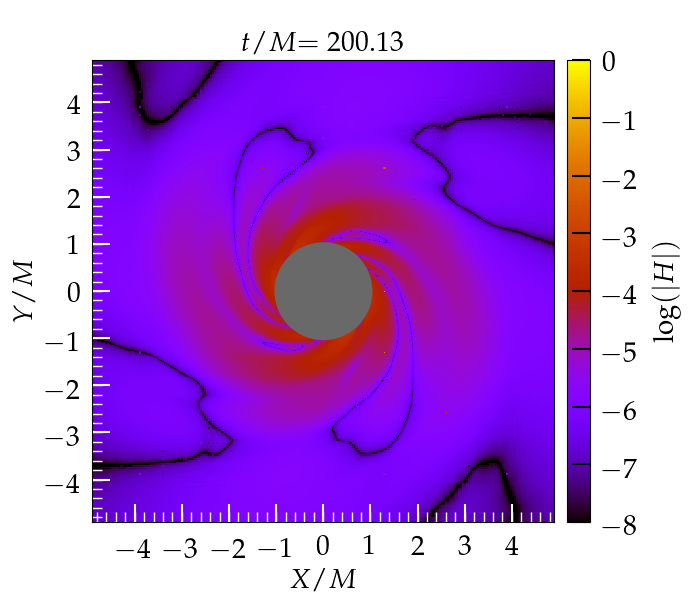}                            
\caption{Hamiltonian violations on the xy-plane at three different instances during the
evolution of the NSNS system. At $t=0$ white dashed curves correspond to the surface of the 
stars.}
\label{fig:ham}                                                                 
\end{center}                                                                     
\end{figure*}

As argued in Ref. \cite{Haensel1989,Koranda1997799} the most compact configuration
will result from a combination of a cold, soft EoS at small densities and a cold stiff one at 
larger ones. Given our current ignorance, there is much freedom in constructing such an allowed EoS.
Doing so
involves choices about the small density part as well as the matching density
to the stiff section. In this work we employ the EoS adopted in Ref. 
\cite{Tsokaros:2019mlz} which we called ALF2cc. 
Eq. (1) relates the pressure to the total energy density when $\GR_0\geq\GR_{0s}$, 
while for $\GR_0\leq\GR_{0s}$ one typically specifies  the pressure in terms of the rest-mass density
in the form of a polytropic-like function. To express the EoS
we can integrate the first law of the thermodynamics in the form
$d\GR/(\GR+P)=d\GR_0/\GR_0$ and 
with the use of Eq. (1) we can write the pressure, energy density and enthalpy
as a function of the rest-mass density. For $\GR_0\geq\GR_{0s}$
\begin{eqnarray}
P   & = & \frac{1}{\GS+1}(\GS\GK \GR_0^{\GS+1} + P_s - \GS\GR_s) \, , \label{eq:precc}\\
\GR & = & \frac{1}{\GS+1}(\GK \GR_0^{\GS+1} + \GS\GR_s - P_s)    \, , \label{eq:rhocc}\\
h   & = & \GK \GR_0^\GS  \, ,  \label{eq:hcc} 
\end{eqnarray}
where $\GK=h_s/\GR_{0s}^\GS$. The value $h_s$ can be evaluated from the 
polytropic piece before the core. 
Spherical equilibrium models for both the ALF2 and ALF2cc EoSs are presented in Fig.
\ref{fig:ALF2cc_TOV} where the gravitational mass $M$ and the compactness of the
stars $M/R$ are plotted vs the rest-mass density. For the ALF2 EoS the maximum
mass is $M_{\rm max}^{\rm TOV}=2.0M_\odot$ at rest-mass density
$1.63\times 10^{15}\ {\rm gr/cm^3}$ and compactness $0.26$. 
For the ALF2cc EoS we have  $M_{\rm max}^{\rm TOV}=4.06M_\odot$
at $\GR_0=6.0\times 10^{14}\ {\rm gr/cm^3}$ and compactness $0.349$.

\begin{figure}                                                                   
\begin{center}                                                                   
\includegraphics[width=0.99\columnwidth]{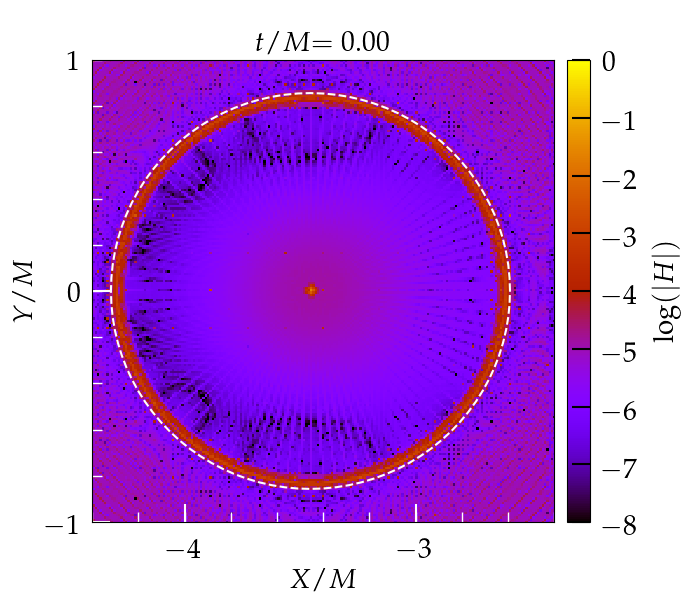}                            
\caption{Zoom in around one NS of the left panel in Fig. \ref{fig:ham} at $t=0$. White
dashed line is the initial data surface of the star. The disk inside the white dashed
line where the violations are large corresponds to the ALF2 crust.}
\label{fig:ham0}                                                                 
\end{center}                                                                     
\end{figure} 

\begin{figure}                            
\begin{center}                                                                   
\includegraphics[width=0.99\columnwidth]{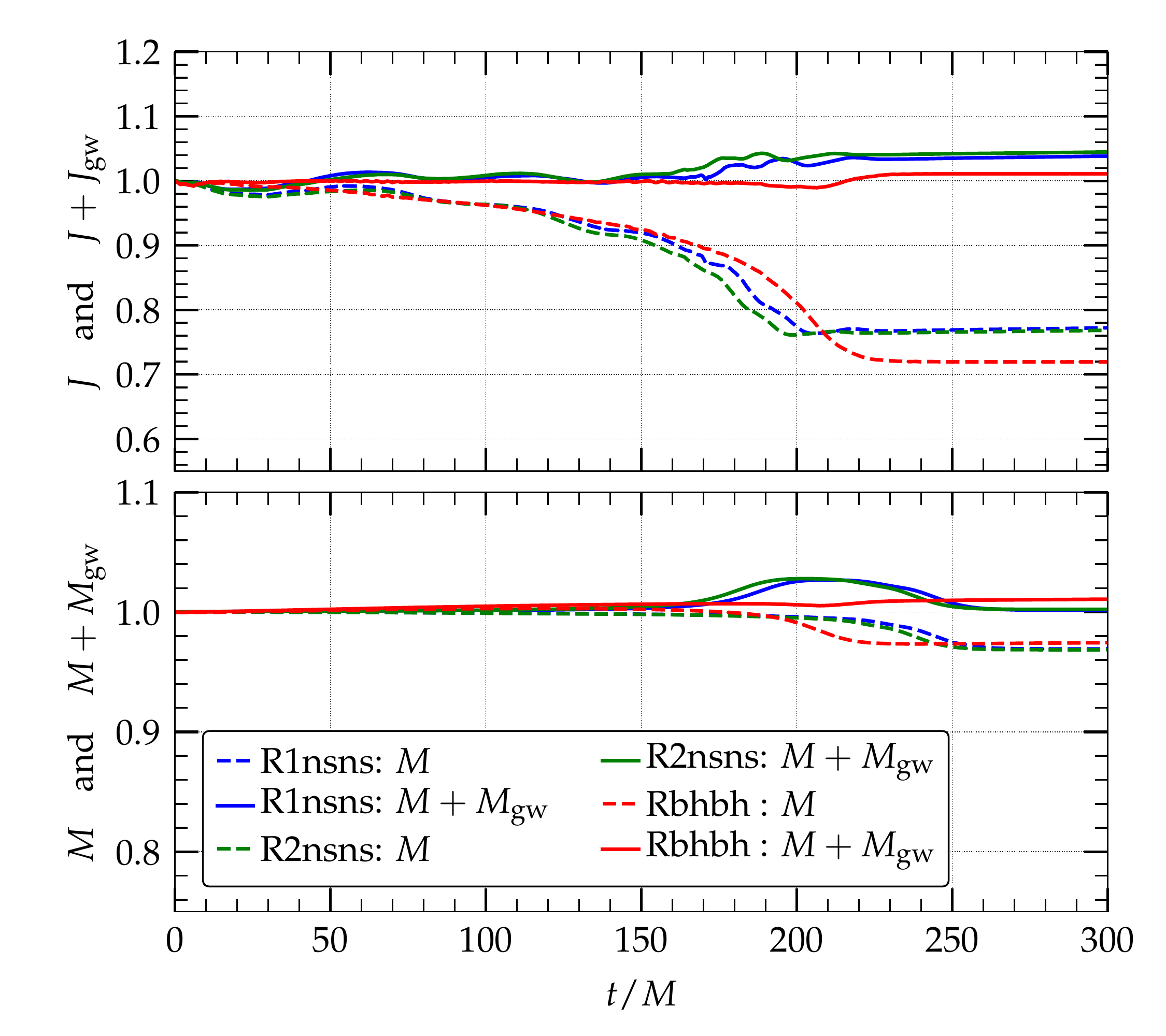}                            
\caption{Conservation of normalized angular momentum $J/J(0)$
and mass $M/M(0)$ for all our binary systems. 
Solid lines are the total angular momentum/mass that includes the contribution carried off by GWs, 
while dashed lines show the angular momentum/mass inside our computational domain.  }  
\label{fig:jmgw}                                                                 
\end{center}                                                                     
\end{figure}

\section{Apparent horizon area and ringdown frequencies}

One key characteristic of the stationary black holes at the end of our simulations
both for the NSNS and BHBH runs is their horizon area. We compute the area in two ways:
first as a surface integral over the apparent horizon, and second by using the Kerr formula
$A=4\pi(r_{+}^2+a^2)=8\pi M_{\rm bh}r_{+}$, where 
$r_{+}=M_{\rm bh}+\sqrt{M_{\rm bh}^2-a^2}$ is the 
Boyer-Lindquist radial coordinate of the event horizon. For the latter case we use 
$a=J_{\rm bh}/M_{\rm bh}^2$, where $J_{\rm bh}$ and $M_{\rm bh}$ are the BH angular
momentum and mass, respectively, as calculated by the isolated horizon formalism \cite{dkss03}.
For both the NSNS and the BHBH runs the two diagnostics yield almost identical results,
which shows that (1) the apparent horizon coincides with the event horizon and (2)
the area is indeed the one given by the Kerr metric. Comparing the 
areas of the remnant BHs of the NSNS and BHBH runs we find that they agree to 
$\sim 3\%$ (though their coordinate shapes are quite different). 

In Fig. \ref{fig:ringdown} we plot the dominant modes $M_{\rm bh}\GO_{22}$ (blue curve) and  
$M_{\rm bh}\GO_{44}$ (red curve) discussed in the text. Vertical
solid lines correspond to the retarded times at which the amplitudes of $h^{22}$ (solid blue line) 
and $h^{44}$ (solid red line) acquire their maximum values. The vertical dashed lines 
correspond to $\Delta t_{\rm ret}=10M_{\rm bh}$ after peak strain times, at which time
we assume that the ringdown starts.
Black curves are best fits for the (4,4) modes and horizontal lines correspond to average
asymptotic values. The frequencies of the (2,2) modes for the NSNS and BHBH binaries are very 
close to each other and lead to determination of the dimensionless BH spin that is consistent
with the values presented in Fig. 4 and discussed in the previous paragraphs.

\section{Diagnostics}

In Fig. \ref{fig:ham} we plot the violations of the Hamiltonian constraints
on the xy plane at $t=0$, and just before and after merger. For the initial data,
the general level is very similar to the violations presented in 
Ref. \cite{Tsokaros:2016eik} where a simple $\Gamma=2$ polytropic EoS was used. 
In the $t=0$ panel white dashed curves (almost circles) correspond to the surfaces of the
stars as calculated from the \cocal code.
The violations of the constraints at the center of the star,
which have been seen and explained in Ref. \cite{Tsokaros:2016eik}, are due to the
spherical coordinate system of the \cocal code, and are washed out almost immediately 
after the evolution starts.
The red circles where the violations are also larger
are in reality thin shells that can be seen clearly in Fig. \ref{fig:ham0}, where we zoom into
the NS on the negative x axis. The width of the shell corresponds to the ALF2 crust
and is the region where the rest-mass density drops steeply from a finite value to zero
(see Fig. 1).
As the binary inspirals these violations similarly track the paths of the stars and 
contaminate the region around them (middle panel in Fig. \ref{fig:ham}). With the formation of the AH these
large violations progressively move inside it and disappear.

In the top panel of Fig. \ref{fig:jmgw} we show the normalized angular momentum 
$J/J(0)$ as calculated by a 
surface integral close to the outer boundary (dashed lines) as well as the total 
angular momentum of the systems which includes the one emitted by GWs, $(J+J_{\rm gw})/J(0)$.
The BHBH binary conserves angular momentum better than $2\%$ while the NSNS binary better than 
$5\%$. At the end of the simulations $23\%$ of the total angular momentum of the NSNS systems has
been radiated away while for the BHBH binary that number is $28\%$. This is also consistent
with the spin angular momemntum diagnostics of the remnant refered to Fig. 4.
Similarly, conservation of mass is shown in the bottom panel of Fig. \ref{fig:jmgw}, with the 
radiated mass being  $3\%\approx 0.24 M_\odot$ of the total.

\section{Phase extrapolation}
In order to compute the phase difference at infinite resolution between the NSNS and BHBH binaries 
we note that if $g_1,\ g_2$ are the quantities at two resolutions $h_1,\ h_2$, then 
\be
g_\infty \approx g_1 + (g_1 - g_2)\frac{1}{(h_2/h_1)^k -1}  \  ,
\label{eq:ginf}
\ee
where $k$ is the order of convergence. In our NSNS case $h_2/h_1=1.2$ and the order of convergence
of the \illinois code is $k=2$ \cite{Etienne:2007jg,els10,Tsokaros:2019anx,Espino:2019xcl}.
Using $\Delta\Phi_{22}$ at resolutions R1nsns and R2nsns (Fig. 3, bottom panel)
and Eq. (\ref{eq:ginf}), one can calculate $\Delta\Phi_{22}$ at infinite resolution as a function of 
the frequency. For a select set of frequencies $f_{22}=700,\ 800,\ 1000$ Hz and $k=2$ we find
$\Delta\Phi_{22}\sim 5$ rad. Note that a smaller order of convergence will increase this phase
difference and thus predict higher probability for distinguishability.

\end{document}